\documentclass[12pt,a4paper]{article}

\usepackage{subfigure}
\usepackage{array}
\usepackage{graphicx}
\usepackage{amsmath}      
\usepackage{amssymb}
\usepackage{mathrsfs}       
\usepackage{bm}            
\usepackage{slashed}
\usepackage{threeparttable}
\usepackage[colorlinks, linkcolor=black, anchorcolor=black, citecolor=black]{hyperref}
\usepackage[amsmath,thmmarks,hyperref]{ntheorem}
\usepackage{soul}

\title{The Electromagnetic Decays of $B^{\pm}_c(2S)$}
\author{Tianhong Wang\footnote{thwang@hit.edu.cn},~Yue Jiang,~Wan-Li Ju,~Han Yuan\\and~~Guo-Li Wang\footnote{gl\_wang@hit.edu.cn}\\
{\it \small   Department of Physics, Harbin Institute of Technology,
Harbin, 150001, China} }
\date{\today}

\begin{document}
\maketitle

\begin{abstract}
We calculate the electromagnetic (EM) decay widths of the $B^{\pm}_c(2S)$ meson, which is observed recently by the ATLAS Collaboration. The main EM decay channels of this particle are $1{^3S_1}\gamma$ and $1{P}\gamma$, which, in literature, are estimated to have the branching ratio of about $1/10$. In this work, we get the partial decay widths: $\Gamma(2{^1S_0}\rightarrow 1{^3S_1}\gamma)=0.192$ keV, $\Gamma(2{^1S_0}\rightarrow 1{P_1}\gamma) = 2.24$ keV and $\Gamma(2{^1S_0}\rightarrow 1{P_1^\prime}\gamma) = 11.4$ keV. In the calculation, the instantaneous approximated Bethe-Salpeter method is used. For the $P$-wave $B_c$ mesons, the wave functions are given by mixing the $^3P_1$ and $^1P_1$ states. Within the Mandelstam formalism, the decay amplitude is given, which includes the relativistic corrections. 
\end{abstract}

\section{Introduction}

Since its discovery by the CDF~\cite{CDF}, the $B_c$ meson has attracted lots of attentions. The reason is that this particle is the only heavy meson consisting of two heavy quarks with different flavors which forbids the annihilation decays to photon and gluon. For this unique nature, the ground state of $B_c$ meson which is below the $BD$ threshold, can only decay through the weak interaction. As a result, it has a very long lifetime which is about $10^{-12}$ s~\cite{PDG}. This provides an ideal platform to study weak decays~\cite{Chang,Nobes, Cho} and even some new physics beyond the Standard Model~\cite{bsm1,bsm2,bsm3,bsm4}.

But the ground state $B_c$ meson is not the whole story. As is known to all, quark potential models predict rich heavy meson spectra. In experiments, lots of these particles have been found, especially at the charmonium and bottomonium sectors. As for the heavy-light mesons, such as $D$, $D_s$, $B$ and $B_s$, the corresponding $P$ wave states have also been found, while for the $B_c$ meson, only the ground state shows itself, until very recently the ATLAS Collaboration at the LHC found the excited $B_c$ meson~\cite{atlas}. This particle is detected through the decay channel: $B_c^{\pm}(2S)\rightarrow B_c^{\pm}(1S)\pi^+\pi^-$ by using $4.9~{\rm fb^{-1}}$ of 7 TeV and $19.2~{\rm fb^{-1}}$ of 8 TeV $pp$ collision data, which gives the mass $6842\pm 4\pm 5$ MeV. 

For its mass is also below the $BD$ threshold, the $B_c(2S)$ meson cannot decay through the OZI-allowed channels. Except the soft gluon radiation process (two pion channel), which has the partial width of $\sim 50$ keV~\cite{Ei} ($\sim 64$ keV~\cite{Ke}) theoretically, it can also transit to the lower states by the electromagnetic decays (in Ref.~\cite{Ei}, the branching ratio of EM channels is larger than $9\%$, which is considerable). Because EM decay channels are more clear compared the strong decay ones, they are important to study the inner structure of particles. Additionally, other $B_c$ mesons could be found through EM decays. For example, the $B_c^\ast(1S)$ state could be detected through the channel: $B_c(2S)\rightarrow B_c(1S)\gamma$ or $B_c^\ast(1S)\rightarrow B_c(1S)\gamma$~\cite{WZG}. Similarly one can also search the $P$ wave states through one photon decay channels. This is important for the study of $B_c$ spectra. Nowadays, there are not enough data collected to reveal more properties of this particle, e.g. the total decay width and branching ratio of specific decay channels. Fortunately, the LHC has started the second running turn. With more data collected, we could hopefully get more informations about this particle, and also, maybe other excited states will show themselves. 

In this paper we get the charm-beauty spectroscopy and corresponding wave functions by using the Bethe-Salpeter (BS) equation~\cite{BS1}. It is a relativistic equation to describe two-body bound states. To solve this equation, we use the instantaneous approximation to the interaction kernel, which results in the three-dimensional Salpeter equation~\cite{BS2}. By constructing the appropriate form of the wave functions for the charm-beauty mesons with different spin-parity, we could get the eigenvalue equations. For the interaction kernel, the screened Cornell potential is applied. This model could describe most of known heavy mesons, especially for those whose masses are below thresholds. As for those which have OZI-allowed decay channels, the predicted mass value usually has a deviation of hundreds of MeV to the experimental result. This can be explained by considering the coupled-channel effects. Here for the charm-beauty system, the ground states of $S$ and $P$ waves are under the $BD$ threshold, which means we could safely use the potential model. 

The paper is organized as follows. In the next section, we briefly outline the theoretical formalism. The wave functions for different mesons are constructed, and EM transition amplitude are given within the Mandelstam formalism. In section 3, we present our results and compare them with those of other models. Finally we draw the conclusions.

\section{Theoretical calculations}

Two-body bound states are well described by the Bethe-Salpeter equation, which has the following form
\begin{equation}\label{BS}
\begin{aligned}
S^{-1}_1(p_1)\chi_{_P}(q)S^{-1}_2(-p_2)=i\int\frac{d^4k}{(2\pi)^4} V(P; q,k)\chi_{_P}(k),
\end{aligned}
\end{equation}
where $\chi_{_P}(q)$ is the BS wave function and $V(P; q,k)$ is the interaction kernel. $P$ and $q$ are the total momentum and relative momentum, respectively. $p_1$ and $p_2$ are the momenta of quark and antiquark, respectively. They are related by the following relation
\begin{equation}
 p_{i} = \frac{m_{i}}{m_{1}+m_{2}}P+Jq.
\end{equation}
where $J=(-1)^{i+1}$ for $i=1,2$.

We decompose the momenta into two parts by projecting to the meson momentum $P$ just as Ref.~\cite{Kim} did, which means $p_{i}^\mu \equiv p_{i_P} \frac{P^\mu}{M} + p_\perp^\mu$, where $p_{i_P}\equiv\frac{P\cdot p_i}{M}$. By defining $\omega_i=\sqrt{m_i^2-p_{i\perp}^2}$, we can introduce the projection operator
\begin{equation}
\Lambda_i^{\pm}(p^\mu_{i\perp})\equiv\frac{1}{2\omega_i}[\frac{\slashed P}{M}\omega_i\pm(\slashed p_{i\perp} + J m_i)],
\end{equation}
with which, the quark (antiquark) propagator $S_i(Jp_i)$ can be rewritten as
\begin{equation}
\label{propagator}
\begin{aligned}
&-iJS_i(Jp_i)=\frac{\Lambda^+_i}{p_{i_P}-\omega_i+i\epsilon}+\frac{\Lambda_i^-}{p_{i_P}+\omega_i-i\epsilon}.
\end{aligned}
\end{equation}

The transition amplitude (see Fig.~1) can be written as 
\begin{equation}
\begin{aligned}
T^\xi = (2\pi)^4\delta^4(P-P_f - k) \mathcal M^\xi.
\end{aligned}
\end{equation}
Within the Mandelstam formalism~\cite{Man}, the Feynman amplitude has the form
\begin{equation}
\begin{aligned}
\mathcal M^\xi&=\int\frac{d^4q}{(2\pi)^4}\frac{d^4q_f}{(2\pi)^4}{\rm Tr}\left[\overline\chi_{_{P_f}}(q_f)Q_1e\gamma^\xi\chi_{_P}(q)(2\pi)^4\delta^4(p_2-p_{f2})S_2^{-1}(-p_2)\right.\\
&\left.~~~~+\overline\chi_{_{P_f}}(q_f)(2\pi)^4\delta^4(p_1-p_{f1})S_1^{-1}(p_1)\chi_{_P}(q)Q_2e\gamma^\xi\right]\\
&=\int\frac{d^4q}{(2\pi)^4}\frac{d^4q_f}{(2\pi)^4}{\rm Tr}\left[S_2(-p_{f2})\overline\eta_{_{P_f}}(q_f)S_1(p_{f2})Q_1e\gamma^\xi S_1(p_1)\eta_{_{P}}(q)S_2(-p_2)\right.\\
&\left.~~~~\times(2\pi)^4\delta^4(p_2-p_{f2})S_2^{-1}(-p_2)+S_2(-p_{f2})\overline\eta_{_{P_f}}(q_f)S_1(p_{f2})(2\pi)^4\right.\\
&~~~~\times\delta^4(p_1-p_{f1})S_1^{-1}(p_1)S_1(p_1)\eta_{_P}(q)S_2(-p_2)Q_2e\gamma^\xi\Big],
\end{aligned}
\end{equation}
where $\bar\chi_{_P}(q)$ is defined as $\gamma^0\chi_{_P}^\dagger(q)\gamma^0$; $q_f$ is the relative momentum of the final meson; $Q_1$ and $Q_2$ are the electric charges (in unit of $e$) of two quarks. In the second equation we used the BS equation
\begin{equation}
\chi_{_P}(q)=S_1(p_1)\eta_{_P}(q_\perp)S_2(-p_2),
\end{equation}
where
\begin{equation}
\label{eta}
\eta_{_P}(q_\perp)=\int\frac{d^3\vec k}{(2\pi)^3} V(q_\perp,k_\perp)\varphi_{_P}(k_\perp).
\end{equation}
In the above equation we have used the definition $\varphi(q^\mu_\perp)\equiv i\int\frac{dq_{_P}}{2\pi}\chi_{_P}(q)$ and the  instantaneous approximation $V(P; q, k) \approx V(P; q_\perp, k_\perp)$.

In the next step we will integrate out $q_f$, which is easy by considering the $\delta$ functions in the integrand. To make the calculation simple, we only consider the positive part of wave functions which give the main contribution. As a result, the amplitude has the following form
\begin{equation}
\label{amp}
\begin{aligned}
&\mathcal M^\xi\approx \int\frac{d^4 q}{(2\pi)^4}{\rm Tr}\Bigg[\frac{\Lambda^+_{f2}(q_{_{f\perp}})\bar\eta_{_{P_f}}(q_{_{f\perp}})\Lambda^+_{f1}(q_{_{f\perp}})Q_1e\gamma^\xi\Lambda^+_1(q_\perp)\eta_P(q_\perp)\Lambda^+_2(q_\perp)\frac{\slashed P}{M}}{(-q_P+\frac{P_P}{2}-\omega_{f2}+i\epsilon)(q_P-\frac{P_P}{2}+P_{fP}-\omega_{f1}+i\epsilon)(q_P+\frac{P_P}{2}-\omega_1+i\epsilon)}\\
&~~~~~~~+\frac{\Lambda^+_{f2}(q_{_{f\perp}})\bar\eta_{_{P_f}}(q_{f\perp})\Lambda^+_{f1}(q_{_{f\perp}})\frac{\slashed P}{M}\Lambda^+_1(q_\perp)\eta_P(q_\perp)\Lambda^+_2(q_\perp)Q_2e\gamma^\xi}{(-q_P-\frac{P_P}{2}+P_{fP}-\omega_{f2}+i\epsilon)(q_P+\frac{P_P}{2}-\omega_{f1}+i\epsilon)(-q_P+\frac{P_P}{2}-\omega_2+i\epsilon)}\Bigg]\\
&~~~=\int\frac{d^3\vec q}{(2\pi)^3}{\rm Tr}\left[Q_1e\frac{\slashed P}{M}\overline{\varphi_{P_f}^{++}}(q_\perp+\alpha_2P_{f\perp})\gamma^\xi\varphi_P^{++}(q_\perp)-Q_2e\overline{\varphi_{P_f}^{++}}(q_\perp-\alpha_1P_{f\perp})\frac{\slashed P}{M}\varphi_P^{++}(q_\perp)\gamma^\xi\right]
\end{aligned}
\end{equation}
where in the first step Eq.~(\ref{propagator}) was used. By using residue theorem, we can integrate out $q^0$, and get the three-dimensional form of the amplitude. Further, by using the Salpeter equation~\cite{Kim}
\begin{equation}
\label{salpeter}
\begin{aligned}
&(M-\omega_1-\omega_2)\varphi^{++}_{_P}(q_\perp) = \Lambda_1^+\eta_{_P}(q_\perp)\Lambda_2^+,
\end{aligned}
\end{equation}
($\varphi^{++}$ is defined as $\Lambda^+_1\frac{\slashed P}{M}\varphi\frac{\slashed P}{M}\Lambda^+_2$) we finish the second step in Eq.~(\ref{amp}).

The wave functions for heavy mesons with different spin-parity quantum number were constructed in our former works~\cite{Kim, WGL, WTH}. By solving corresponding Salpeter equations we can get their numerical results. For $0^-$ and $1^-$ states, their wave functions have the following forms
\begin{equation}
\label{0-wave}
\begin{aligned}
\varphi_{_{0^-}}=M\left(f_1\frac{\slashed P}{M}+f_2+f_3\frac{\slashed q_{\perp}}{M}+f_4\frac{\slashed P\slashed q_{\perp}}{M^2}\right)\gamma^5,
\end{aligned}
\end{equation}
\begin{equation}
\label{1-wave}
\begin{aligned}
\varphi_{_{1^-}}&=q_{_{f\perp}}\cdot\epsilon\left(g_1+g_2\frac{\slashed P_f}{M_f}+g_3\frac{\slashed q_{f\perp}}{M_f}+g_4\frac{\slashed P_f\slashed q_{f\perp}}{M_f^2}\right)+M_f{\slashed\epsilon}\left(g_5+g_6\frac{\slashed P_f}{M_f}\right)\\
&~~~~+\left(\slashed q_{f\perp}\slashed\epsilon - q_{_{f\perp}}\cdot\epsilon\right)g_7+\frac{1}{M_f}\left(\slashed P_f\slashed\epsilon\slashed q_{f\perp} - \slashed P_fq_{_{f\perp}}\cdot\epsilon\right)g_8,
\end{aligned}
\end{equation}
where $f_i$s and $g_i$s are functions of $\vec q$ and $\vec q_f$, respectively.
For the $1^+$ and $1^{+\prime}$ states, we mix the wave functions of states with specific charge parity, that is $^3P_1$ and $^1P_1$ states
\begin{equation}
\label{1++wave}
\begin{aligned}
\varphi_{_{^3P_1}}=i\epsilon_{\mu\nu\alpha\beta}\frac{P_f^\nu}{M_f}q_{_{f\perp}}^\alpha\epsilon^\beta\gamma^\mu\left(g_1+g_2\frac{\slashed P_f}{M_f}+g_3\frac{\slashed q_{f\perp}}{M_f}+g_4\frac{\slashed P_f\slashed q_{f\perp}}{M_f^2}\right),
\end{aligned}
\end{equation}
\begin{equation}
\label{1+-wave}
\begin{aligned}
\varphi_{_{^1P_1}}=q_{_{f\perp}}\cdot\epsilon\left(g_1+g_2\frac{\slashed P_f}{M_f}+g_3\frac{\slashed q_{f\perp}}{M_f}+g_4\frac{\slashed P_f\slashed q_{f\perp}}{M_f^2}\right)\gamma_5.
\end{aligned}
\end{equation}

By introducing a mixing angle $\theta_{1P}$, we could get the states corresponding to the physically detected particles
\begin{equation}\label{mixing}
\begin{aligned}
\left(
\begin{array}{c}
|1^{+\prime}\rangle\vspace{0.3cm}\\
|1^{+}\rangle
\end{array}
\right)=
\left(
\begin{array}{cc}
\cos\theta_{1P}& \sin\theta_{1P}\vspace{0.3cm}\\
-\sin\theta_{1P}& \cos\theta_{1P}
\end{array}
\right)
\left(
\begin{array}{c}
|^1P_1\rangle\vspace{0.3cm}\\
|^3P_1\rangle
\end{array}
\right)
\end{aligned}
\end{equation}
To solve the Salpeter equations fulfilled by the wave functions of different states, we use the numerical method. The absolute value of $\vec q$  is discretized and truncated around $4$~GeV where the wave functions $f_i$s and $g_i$s are small enough which shows their convergence is very good. Some details can be found in Refs.~\cite{Kim, WGL}.  Afterwards, we insert Eq.~(\ref{0-wave}) $\sim$ Eq.~(\ref{1+-wave}) into Eq.~(\ref{amp}), and integrate out the relative momentum $q_\perp$.

Here we just give the final results of the decay amplitude. For the process $2{^1S_0}\rightarrow 1{^3S_1}\gamma$, with all the integrals of $q_\perp$ absorbed into the only form factor $t$, we get 
\begin{equation}
\begin{aligned}
\mathcal M^\xi = \epsilon^{\xi\mu\nu\alpha}\epsilon_\mu P_\nu P_{f\alpha} t.
\end{aligned}
\end{equation}
While for $2{^1S_0}\rightarrow 1P_1(1P_1^\prime)\gamma$, there is a polarization vector along with the pseudovector meson, which results in three form factors $s_1\sim s_3$ of the amplitude 
\begin{equation}
\begin{aligned}
\mathcal M^\xi = P\cdot \epsilon P^\xi s_1 + P\cdot \epsilon P_f^\xi s_2 + M^2\epsilon^\xi s_3.
\end{aligned}
\end{equation}

\begin{figure}
\label{feynman}
\centering
\includegraphics[scale=1.0]{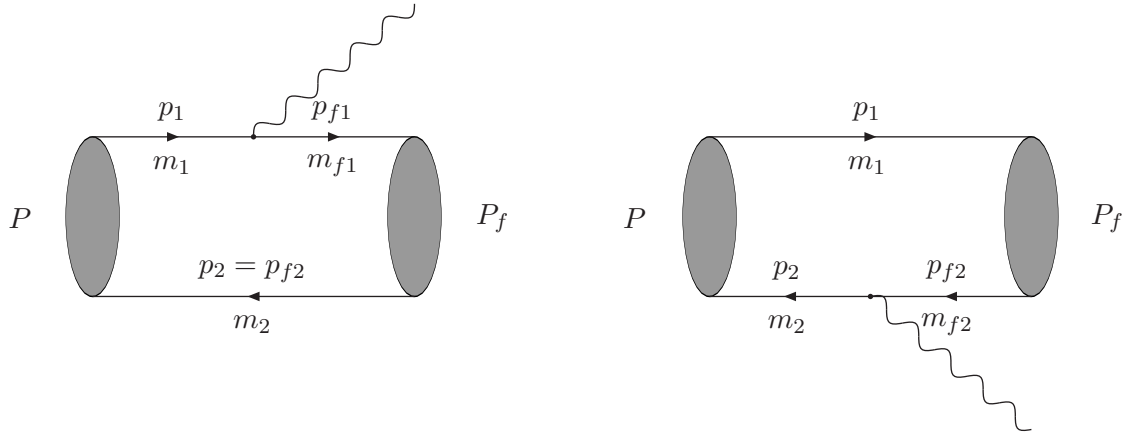}
\caption{Feynman diagrams of the single-photon transition between two heavy mesons. The left diagram represents the photon comes from the quark, while the right one represents the photon comes from the anti-quark.}
\end{figure}

\section{Results and discussions}
To solve the instantaneous BS equation, in Eq.~(\ref{eta}) we used the Cornell potential which includes a linear term and a Coulomb term. In the momentum space it has the following form~\cite{Kim},

\begin{equation}
\begin{aligned}
\label{Cornell}
V({\vec q})=V_s(\vec{q})
+\gamma_0\otimes\gamma^0V_v(\vec{q}),
\end{aligned}
\end{equation}
where
\begin{equation}
\begin{aligned}
&V_{s}(\vec{q})
=-\left(\frac{\lambda}{\alpha}+V_0\right)\delta^{3}(\vec{q})
+\frac{\lambda}{\pi^{2}}\frac{1}{(\vec{q}^{2}+\alpha^{2})^{2}},\\
&~~~~~~~~~~~~~~V_v(\vec{q})=-\frac{2}{3\pi^{2}}
\frac{\alpha_{s}(\vec{q})}{\vec{q}^{2}+\alpha^{2}}.
\end{aligned}
\end{equation}
The running coupling constant is expressed as
\begin{equation}
\begin{aligned}
\alpha_s(\vec{q})=\frac{12\pi}{27}
\frac{1}{{\rm{ln}}\left(a+\frac{\vec{q}^2}{\Lambda_{QCD}}\right)}.
\end{aligned}
\end{equation}
Parameters in above equations have the values: $a=e=2.71828$, $\alpha=0.06$ GeV, $\lambda=0.21$ ${\rm GeV}^2$, $\Lambda_{QCD}=0.27$ GeV, $m_b=4.96$ GeV, $m_c=1.62$ GeV. These parameters were used in Refs.~\cite{WTH, Fu, Ju}, where the reasonable spectra were given. 

In Table~\ref{mass}, we present the masses used in this paper and other models. The experimental value is used for the  mass of $B_c(2S)$. In our model, to get the correct mass spectrum, we adjust the value of $V_0$ in the Cornell potential. For the $B_c(1S)$ meson, whose mass is $6275.6$ GeV, $V_0$ is set to $-0.1837$ GeV. With this parameter fixed, we could get the mass of excited states. For example, it predicts $M_{B_c(2S)}=6.857$ GeV which is about $15$ MeV larger than the experimental central value. So here is a difficulty to make sure all the ground state and excited ones to have correct masses. This phenomenon is common in potential models, especially for the states above thresholds. The potential we used above is too simple (on the one hand for the Coulomb term, only the time-like part is kept; on the other hand the linear term is also an approximation), which we cannot hoped to give very precise results. So here we adjusted the  $V_0$ to be  $-2.05$ GeV  to set the mass of $B_c(2S)$ equals to the experimental value.

Our results for the EM decay of $B_c(2S)$ meson are listed in Table~\ref{Bc2Sff} (form factors) and Table~\ref{Bc2Sdecay} (partial widths). To calculate the $1P\gamma$ and $1P^\prime\gamma$ decay widths, we use the mixed wave functions which are constructed in Eq.~(\ref{mixing}). Here we assume the mixing angle has the value $\theta_{1P}=23.2^\circ$. From Table~\ref{Bc2Sdecay} one can see, for the $B_c^\ast\gamma$ channel, our result is about $2.5$ (5) times smaller than that of Ref.~\cite{Ebert11} (Ref.~\cite{Rai}) while $2$ times larger than those of Refs.~\cite{Ei,Ger}. For Refs.~\cite{Ful, God}, their results are similar to ours. For the $1P\gamma$ channel, ours is very close to the results in Refs.~\cite{Ger, God}. As for the $1P^\prime\gamma$ channel, we get the result $2\sim3$ times larger than those of Refs.~\cite{Ebert03, Ei}. In Refs.~\cite{Ei, Ger, God}, the relativistic effects were included in the potential by adding spin-related interaction Hamiltonians, while the EM transition amplitude was written within the non-relativistic formalism. In Ref.~\cite{Ebert03}, to get the wave functions of mesons, the quasi-potential equation of the Schr$\ddot{\rm o}$dinger type was used. The authors also calculated the relativistic corrections to the transition amplitude by considering the Lorentz transformation of wave functions. In our mode, on the one hand, the wave function fulfills the instantaneous Bethe-Salpeter equation which already includes the relativistic effects (In Ref.~\cite{Hady}, the BS equation is also used, while the transition amplitude has a non-relativistic form just as that of Ref.~\cite{Ei}), and on the other, the transition amplitude we used has a relativistic covariant form. 

In Fig.~2, we plot the partial decay widths of $B_c(2S)$ meson, the mass of which has changed from $6.830$ GeV to $6.855$ GeV. Here we have scaled $\Gamma_{B_c(2S)\rightarrow B_c^\ast\gamma}$ to ten times larger. From the plot we can see that all the three channels have partial decay widths which increase with the initial meson mass. For the $B_c^\ast\gamma$ channel, the partial width is $0.186\sim 0.198$ keV, while for the $1P\gamma$ and $1P^\prime\gamma$, the partial widths are $2.11\sim 2.36$ keV and $10.4\sim 12.3$ keV, respectively. For the last two channels, we have used the mixing angle $\theta_{1P}=23.2^\circ$. One can see, the decay widths do not change very much at this interval of initial meson mass. Considering the statistical and system errors of experiment value, the $B_c(2S)$ meson could have mass within this interval. Our result shows that changing mass of $B_c(2S)$ causes its EM partial decay widths to vary within $10\%$.

In Fig.~3, the partial decay widths of $B_c(2S)$ to $1P$ and $1P^\prime$ which changed with the mixing angle are plotted. Here we use a different condition to that in Fig.~2, that is, the phase space is fixed. For the $1P\gamma$ channel, the partial decay width changes like a sinusoidal function, which means, when the mixing angle is $0$, it has the smallest value, while if the mixing angle is $\pi/2$, it has the largest value. For the $1P^\prime\gamma$ channel, the partial width varies like a cosine function. Here we also gave the decay widths when the mixing angle changed $\pm 10^\circ$, which are represented by the dashed vertical lines. When the $\theta_{1P}=13.2^\circ$, we get $\Gamma_{B_c(2S)\rightarrow 1P\gamma}=0.659$ keV and $\Gamma_{B_c(2S)\rightarrow 1P^\prime\gamma}=12.0$ keV; for $\theta_{1P}=33.3^\circ$, we get $\Gamma_{B_c(2S)\rightarrow 1P\gamma}=4.57$ keV and $\Gamma_{B_c(2S)\rightarrow 1P^\prime\gamma}=9.85$ keV. So when the mixing angle changes within this interval, the partial width of $1P\gamma$ channel varies about $2\sim 3$ times, while that of $1P^\prime\gamma$ channel changed a little. Usually different models used different mixing angles, such as $20.4^\circ$~\cite{Ebert03}, $17.1^\circ$~\cite{Ger} and $33.4^\circ$~\cite{Dav}. Just as Ref.~\cite{God} mentioned, different models could give very different results for the EM transitions  are sensitive to the mixing. The measurement of radiative decays may provide a possible way to distinguish between the different models.

As a conclusion, we have calculated the electromagnetic decays of $B_c(2S)$ meson. Our results are compatible with that of other models. The $B_c(2S)\rightarrow 1P\gamma$ or $1P^\prime\gamma$ channels have larger decay widths than that of $B_c(2S)\rightarrow B_c^\ast(1S)\gamma$. In the former two channels we used the mixing angle $23.2^\circ$, which makes the partial width of $1P^\prime\gamma$ channel about one order larger than that of $1P\gamma$ channel. These results will be useful for the future experiments to study properties of $B_c(2S)$ meson and other excited states.

\begin{figure}
\label{figure2}
\centering
\includegraphics[scale=1.15]{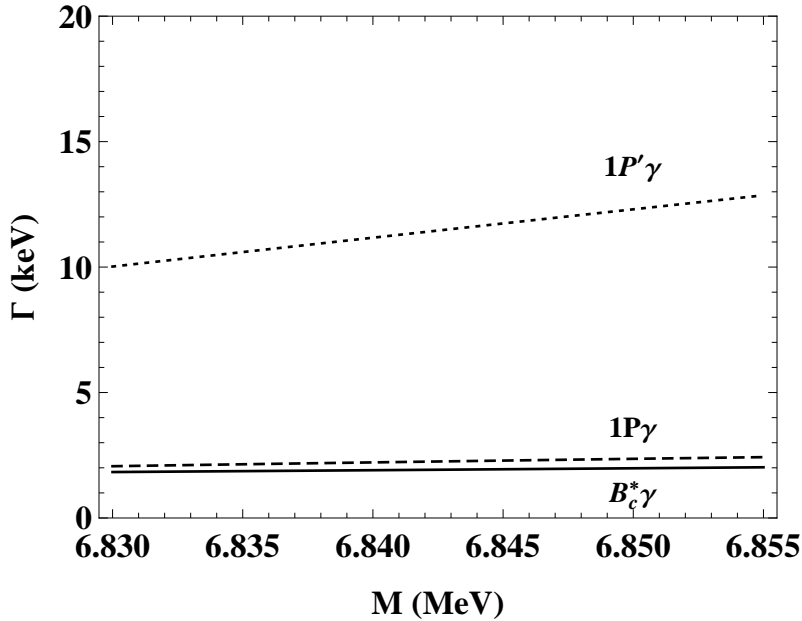}
\caption{The partial decay widths for three decay channels change with the mass of the $B_c(2S)$ state. The dotted line is for the $1P^\prime\gamma$ channel; the dashed line is for the $1P\gamma$ channel; the solid line is for the $B_c^\ast(1S)\gamma$ channel. For the $B_c^\ast(1S)\gamma$ channel, to make the results more clearly we have scaled the value to ten times larger.}
\end{figure}

\begin{figure}
\label{figure3}
\centering
\includegraphics[scale=1.12]{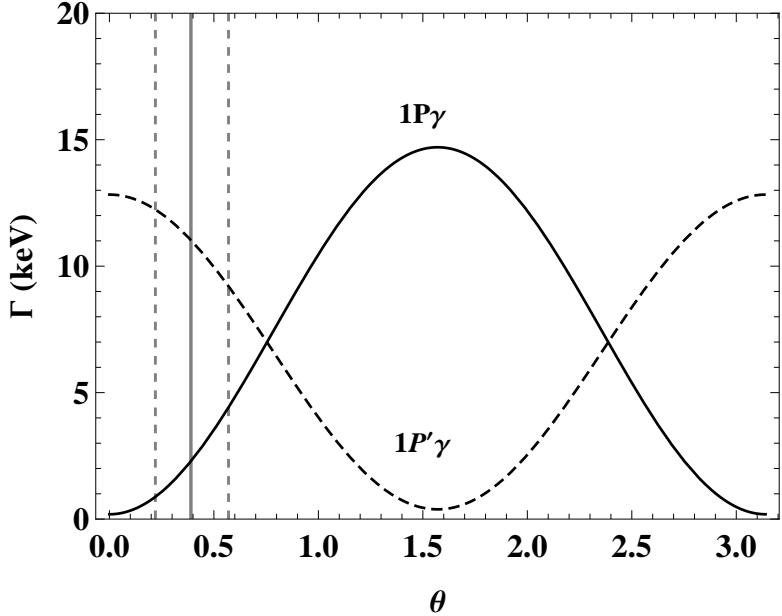}
\caption{The partial decay widths for the $1P\gamma$ (solid line) and $1P^\prime\gamma$ (dashed line) change with the mixing ange (from $0$ to $\pi$). The solid vertical line represents the position where $\theta=0.405~(23.2^\circ)$ is adopted, and the dashed vertical ones are those changed $\pm 10\%$, respectively.}
\end{figure}

\begin{table}
\caption{The masses (MeV) of $B_c$ mesons used in our calculation and other models. In this paper, we use the experimental values for the $B_c(1S, 2S)$ states. For the $B_c^\ast(1S)$ state, we use the same mass as that of Ebert {\sl et al.}}
\label{mass}
\setlength{\tabcolsep}{0.1cm}
\centering
\begin{threeparttable}
\begin{tabular*}{\textwidth}{@{}@{\extracolsep{\fill}}cccccccc}
\hline\hline
State&Ours&Ref.~\cite{Ebert11}\tnote{$\dagger$}&Ref.~\cite{Ei}&Ref.~\cite{Ger}&Ref.~\cite{Ful}&Ref.~\cite{God}&Ref.~\cite{Rai}\\
 \hline
{\phantom{\Large{l}}}\raisebox{+.2cm}{\phantom{\Large{j}}}
$B_c(1S)$&6275.6&6272&6264&6253&6286 &6271&6.278\\ 
{\phantom{\Large{l}}}\raisebox{+.2cm}{\phantom{\Large{j}}}
$B_c(2S)$&6842&6842&6856&6867&6882 &6855&6.863\\ 
{\phantom{\Large{l}}}\raisebox{+.2cm}{\phantom{\Large{j}}}
$B_c^\ast(1S)$&6333&6333&6337&6317&6341 &6338&6.331\\ 
{\phantom{\Large{l}}}\raisebox{+.2cm}{\phantom{\Large{j}}}
$1^+(1P)$&6732&6743&6730&6717&6737 &6741&6.767\\
{\phantom{\Large{l}}}\raisebox{+.2cm}{\phantom{\Large{j}}}
$1^{+\prime}(1P)$&6749&6750&6736&6729&6760 &6750&6.769\\
\hline\hline
\end{tabular*}
	\begin{tablenotes}
	\item[$\dagger$] In Ref.~\cite{Ebert03}, the authors got $M_{B_c(2S)}=6.835$ MeV.
	\end{tablenotes}
\end{threeparttable}
\end{table}

\begin{table}
\caption{Form factors (${\rm GeV}^{-1}$) for three EM decay channels of $B_c(2S)$. The errors come from the variation of parameters in the Cornell potional by $\pm5\%$.}
\label{Bc2Sff}
\setlength{\tabcolsep}{0.1cm}
\centering
\begin{tabular*}{\textwidth}{@{}@{\extracolsep{\fill}}ccccc}
\hline\hline
form factors&$t$&$s_1$&$s_2$&$s_3$\\ \hline
{\phantom{\Large{l}}}\raisebox{+.2cm}{\phantom{\Large{j}}}
$B_c(2S)\rightarrow B_c^\ast(1S)\gamma$&$1.50^{+0.43}_{-0.37}\times 10^{-2}$&0&0&0 \\ 
{\phantom{\Large{l}}}\raisebox{+.2cm}{\phantom{\Large{j}}}
$B_c(2S)\rightarrow 1^+(1P)\gamma$&0&$0.970^{+0.102}_{-0.094}$&$-0.482^{+0.078}_{-0.087}$&$-7.75^{+0.47}_{-0.51}\times 10^{-3}$\\ 
{\phantom{\Large{l}}}\raisebox{+.2cm}{\phantom{\Large{j}}}
$B_c(2S)\rightarrow 1^{+\prime}(1P)\gamma$&0&$1.12^{+0.20}_{-0.17}$&$-2.54^{+0.24}_{-0.26}$&$1.90^{+0.12}_{-0.12}\times 10^{-2}$\\ 
\hline\hline
\end{tabular*}
\end{table}

\begin{table}
\caption{Partial widths (keV) for three EM decay channels of $B_c(2S)$. The errors come from the variation of parameters in the Cornell potional by $\pm5\%$.}
\label{Bc2Sdecay}
\setlength{\tabcolsep}{0.05cm}
\centering
\begin{tabular*}{\textwidth}{@{}@{\extracolsep{\fill}}cccccccc}
\hline\hline
Channel&Ours&Ref.~\cite{Ebert03}&Ref.~\cite{Ei}&Ref.~\cite{Ger}&Ref.~\cite{Ful}&Ref.~\cite{God}&Ref.~\cite{Rai}\\ \hline
{\phantom{\Large{l}}}\raisebox{+.2cm}{\phantom{\Large{j}}}
$B_c(2S)\rightarrow B_c^\ast(1S)\gamma$&$0.192^{+0.083}_{-0.062}$&0.488&0.093&0.096&0.139&0.3 &$\sim 1$\\ 
{\phantom{\Large{l}}}\raisebox{+.2cm}{\phantom{\Large{j}}}
$B_c(2S)\rightarrow1^{+}(1P)\gamma$&$2.24^{+0.30}_{-0.27}$&1.02&0.0&1.9&6.4&1.3&0.62\\ 
{\phantom{\Large{l}}}\raisebox{+.2cm}{\phantom{\Large{j}}}
$B_c(2S)\rightarrow1^{+\prime}(1P)\gamma$&$11.4^{+1.5}_{-1.4}$&3.72&5.2&15.9&13.1&6.1&3.03\\ 
\hline\hline
\end{tabular*}
\end{table}

\section{Acknowledgments}
This work was supported in part by the National Natural Science
Foundation of China (NSFC) under Grant No.~11575048, No.~11405037 and No.~11505039. T.Wang was also supported by PIRS of HIT No.B201506.

\end{document}